# Acoustic realization of quadrupole topological insulators

Yajuan Qi[1], Chunyin Qiu[1*], Meng Xiao[1*], Hailong He[1], Manzhu Ke[1], and Zhengyou Liu[1,2]

[1]Key Laboratory of Artificial Micro- and Nano-Structures of Ministry of Education and School of Physics and Technology, Wuhan University, Wuhan 430072, China

[2]Institute for Advanced Studies, Wuhan University, Wuhan 430072, China

**Abstract**: A quadrupole topological insulator, being one higher-order topological insulator with nontrivial quadrupole quantization, has been intensely investigated very recently. However, the tight-binding model proposed for such emergent topological insulators demands both positive and negative hopping coefficients, which imposes an obstacle in practical realizations. Here we introduce a feasible approach to design the sign of hopping in acoustics, and construct the first acoustic quadrupole topological insulator that stringently emulates the tight-binding model. The inherent hierarchy quadrupole topology has been experimentally confirmed by detecting the acoustic responses at the bulk, edge and corner of the sample. Potential applications can be anticipated for the topologically robust in-gap states, such as acoustic sensing and energy trapping.

[*]Corresponding authors: cyqiu@whu.edu.cn, phmxiao@whu.edu.cn

*Introduction.* The past decade has witnessed an explosive development of topological states of matter in classical wave systems [1-5]. The robustness of the topological systems against disorders and the associated one-way edge states provide new opportunities for manipulating classical waves. Besides the analogues of conventional topological insulators [6-24] and topological semimetals [25-34], the recent interest on higher-order topological insulators (HOTIs) [35,36] has opened a new direction of topological phases in classical systems [37-53]. Different from the conventional topological insulators, in which the *D*-dimensional bulk is gapped and the topological invariant counts the number of gapless modes hosted on the (*D*-1)-dimensional boundaries of the sample, the HOTIs are a new family of topological phases of matter that goes beyond the conventional bulk-boundary correspondence principle. For example, unlike conventional two-dimensional (2D) topological insulators, a 2D second-order HOTI does not exhibit gapless one-dimensional (1D) topological edge states, but instead has topologically protected zero-dimensional (0D) corner states. Due to the flexibility in sample design, HOTIs have been soon implemented in mechanical [37,45], photonic [38,40-44], electrical circuits [39,46,47] and acoustic [48-53] systems.

A quadrupole topological insulator (QTI) [35,36], featuring nontrivial quadrupole moment, has drew extensive attention recently [37-39,41]. In this novel 2D topological system, a bulk quadrupole moment in a finite-sized sample gives rise to surface dipole moments on its 1D edges and to uncompensated charges on the 0D corners. The former contributes to gapped edge modes and the latter exhibits the presence of nontrivial corner modes, which reflects an exotic hierarchy topological inherent in the QTIs [35,36]. Comparing with the other 2D HOTIs, the corner states of QTIs are stably pinned to the middle of the bulk gap due to the inherent chiral symmetries, and thus persist as long as the bulk band exhibits nontrivial quadrupole moment. Though broad attention received, experimental studies on such 2D QTIs are still rather limited [37-39,41]. The QTI theory proposed with a square-lattice tight-binding (TB) model [35,36] requires a $\pi$ flux per plaquette. This demands both positive and negative hoppings in a spinless and time-reversal invariant classical system, where the hoppings must be real-valued. However, fulfilling this harsh requirement usually involves fine-tuning of the parameters [37,41,54], and hence hinders the experimental implementation of the classical QTIs. In particular, so far there is no experimental progress reported for acoustic QTIs, since there is no feasible route proposed to control the sign of acoustic couplings.

In this Letter, we conceive a simple mechanism to generate both positive and negative hoppings in acoustics. That is, linking acoustic cavities with different connectivity according to the field morphologies of acoustic resonators. After testing the design route of controlling hoppings, we present an experimental realization of the acoustic QTI that stringently fulfills the 2D TB model proposed for QTIs [35,36]. The hierarchy topology of our acoustic QTI has been conclusively identified through detecting the acoustic responses at the bulk, edge and corner. In particular, our acoustic QTI exhibits a big bulk gap and the topological corner states are well separated from the gapped edge and bulk states, which is crucial in practical applications such as creating topologically stable acoustic enhancement. All experimental data agree excellently with our numerical simulations performed with COMSOL Multiphysics.

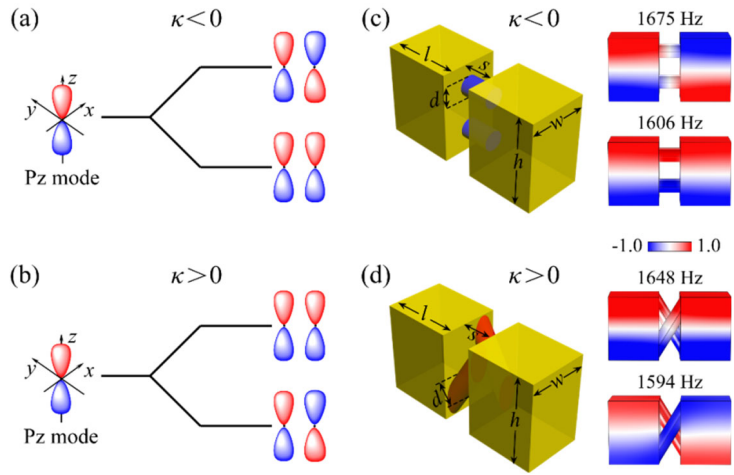

FIG. 1. (a) and (b): Sketches of the split energy levels for two coupled $P_z$ dipoles

with hoppings $\kappa < 0$ and $\kappa > 0$, respectively. For the case of $\kappa < 0$ ($\kappa > 0$), the in-phase coupled mode has an energy lower (higher) than that of out-of-phase one. (c) and (d): Acoustic realizations of the negative and positive hoppings with coupled resonators, respectively. Left: Double cavity structures coupled with different connectivity. The resonance cavities (yellow) and narrow tubes (red or blue) are filled with air and bounded with hard boundaries. Right: Pressure distributions at given eigenfrequencies.

*Design route of positive and negative hoppings in acoustics.* As stated above, a prerequisite for realizing acoustic QTIs is to design positive and negative hoppings independently. So far there is no explicit route conceived to accomplish this goal, despite the fact that TB models have been frequently implemented in acoustics. Here we start from two identical resonators connected with a coupler, associated with effective Hamiltonian

$$H = \begin{pmatrix} \omega_0 & \kappa \\ \kappa & \omega_0 \end{pmatrix}, \qquad (1)$$

in which $\omega_0$ is the frequency of the two resonators, and $\kappa$ represents a real-valued coupling coefficient between them. The composed system exhibits split eigenfrequencies $\omega_\pm = \omega_0 \pm \kappa$ associated with eigenvectors $\frac{1}{\sqrt{2}}(1, \pm 1)^T$: the signs + and − characterize the modes formed by in-phase and out-of-phase couplings, respectively. Therefore, a negative (positive) coupling $\kappa$ suggests that the in-phase coupled mode has an energy lower (higher) than that of out-of-phase one. This is illustrated in Figs. 1(a) and 1(b) by a coupled $P_z$-dipole system. In acoustics, both configurations can be implemented by a pair of identical air cavities coupled with narrow tubes, which are distinguished with straight-linked and cross-linked connectivity, as shown in Figs. 1(c) and 1(d). Physically, the cavity resonators emulate atomic orbitals and the narrow tubes introduce hoppings between them. Here the cavity parameters $l = 8.12$ cm, $w = 6.90$ cm and $h = 10.45$ cm are selected to ensure the frequency of the $P_z$ mode (1627 Hz) far away from the other cavity modes (*Supplemental Materials*). When the air cavities are straightly connected with two identical tubes (of diameter $d = 2.32$ cm and length $s = 3.48$ cm, at a distance of $h/4$ to the top or bottom surface), the composed system exhibits an in-phase coupled $P_z$ mode at a frequency (1606 Hz) lower than that of out-of-phase one (1675 Hz). That is, the straight-linked system has a negative coupling for $P_z$ modes. By contrast, the system exhibits a positive coupling when the cavities are crosswise connected, in which the in-phase coupled mode occurs at a frequency (1648 Hz) higher than that of out-of-phase one (1594 Hz). Note that the hopping strengths depend mostly on the sizes and positions of the narrow tubes, and the central frequencies of the coupled systems may shift from the single cavity resonance due to the introduction of coupling tubes (*Supplemental Materials*).

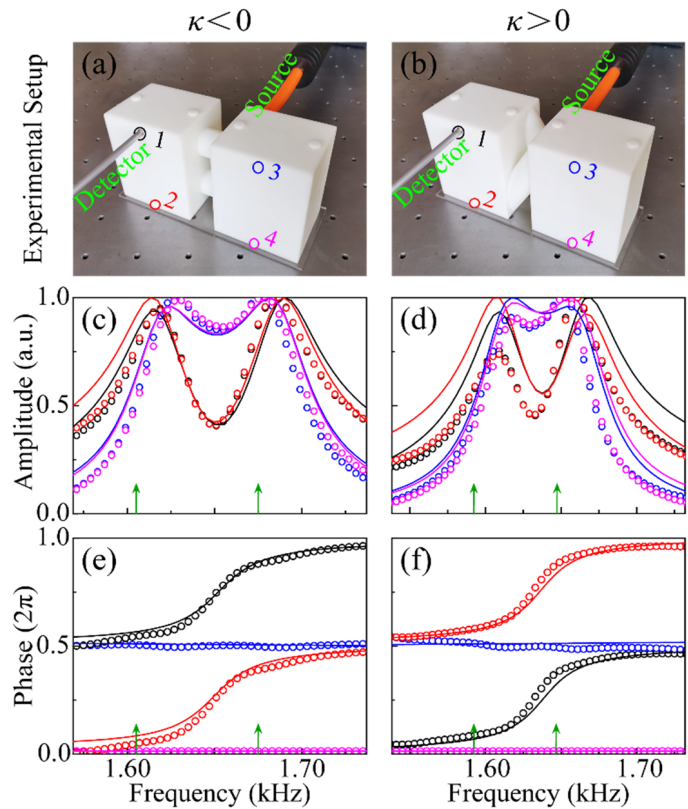

FIG. 2. (a) and (b): Experimental setups for testing the hopping signs of the straight-linked and cross-linked double cavity systems, respectively. The acoustic

signal is injected from the backside of the sample, and four typical positions (labeled with color circles) in the front are used for detection. (c) and (d): Measured (circles) and simulated (lines) pressure amplitude responses at the four locations (plotted with consistent color). The green arrows indicate the eigenfrequencies of the coupled $P_z$ modes. (e) and (f): The associated phase spectra.

*Experimental identification of the sign of the designed hoppings.* The design route for achieving desired hopping sign was confirmed in our airborne sound experiments. Figures 2(a) and 2(b) show the experimental setups. The samples were fabricated precisely via mature 3D-printing technique with photosensitive resin, which can be safely viewed as acoustically rigid for airborne sound. For each case, a broadband point-like sound source, launched from a subwavelength-sized tube, was injected into the right cavity from the backside, and the pressure responses of four typical positions were detected through the holes (of radii ~0.4 cm) perforated in the front of the sample, which were sealed when not in use. Figures 2(c) and 2(d) present the measured amplitude spectra (color circles) for the straight-linked and cross-linked systems, respectively, together with the simulation results (color lines) for comparison. (Each amplitude spectrum was normalized by the measurement without sample and then rescaled to its maximum.) As expected, in each case two resonance peaks emerge near the eigenfrequencies (green arrows) of the coupled $P_z$ modes. Figures 2(e) and 2(f) present the phase spectra under the reference of the position 4, which show clearly a phase difference $\sim \pi$ between the positions 1 and 2 (or 3 and 4), an indication of the $P_z$ modes in both configurations. In particular, for the straight-linked cavity structure [Fig. 1(e)], the phase difference between the positions 1 and 3 (or 2 and 4) approaches zero at the lower frequency while approaches $\pi$ at the higher frequency. This concludes unambiguously a negative hopping since the in-phase coupled mode occurs at the lower frequency than that of out-of-phase one. In contrast, for the cross-linked structure [Fig. 1(f)], the phase responses indicate an in-phase mode at the higher frequency and thus conclude a positive hopping. All the experiments match quantitatively with the simulations. It is worth pointing out that here we take advantage of the unique field profile of the $P_z$ mode to achieve positive and negative hoppings independently. In fact, the cavity also supports $P_x$ dipole at a higher frequency; however, both configurations contribute positive hoppings for $P_x$ mode (*Supplemental Materials*).

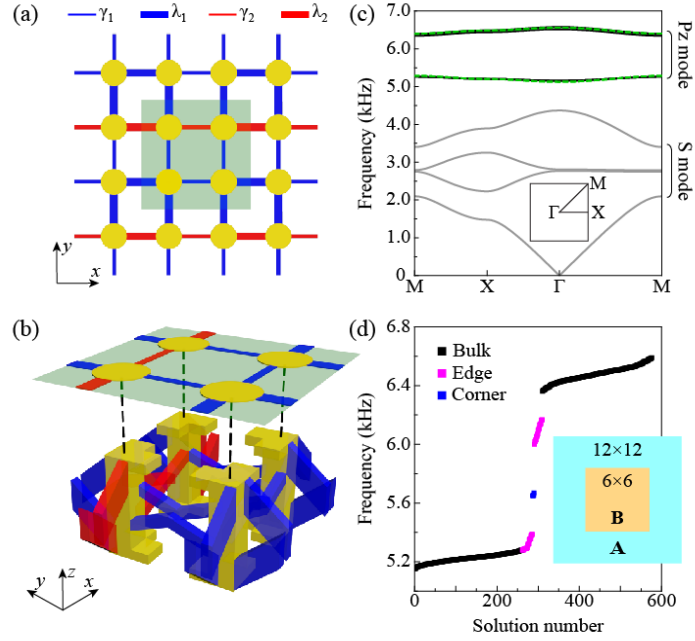

FIG. 3. (a) Square-lattice TB model with positive (red) and negative (blue) nearest-neighbor hoppings $\lambda_{1,2}$ and $\gamma_{1,2}$, where the shaded square sketches a four-site unit cell. This hoping distribution results in a $\pi$ flux per plaquette. (b) Unit structure of our acoustic system that emulates the TB model: yellow for resonant cavities, and blue and red for coupling tubes. (c) Band structure. The $P_z$ bands (black solid lines) are captured well by the TB model (green dashed lines). Inset: the first Brillouin zone of the square lattice. (d) Eigenvalue spectrum for a finite-sized sample constructed with the topologically distinct quadruple phases **A** and **B** (inset), identified with bulk, edge and corner states through inspecting eigenfields.

*Acoustic QTI and topological in-gap states*. Our design starts from the square-lattice

TB model for QTI theory [35,36]. As sketched in Fig. 3(a), each unit cell contains four sublattices: the yellow disks represent lattice sites and the bonds are hoppings between them. For a spinless and time-reversal invariant system, the hoppings can only be real, either positive or negative. Considering the fact that each plaquette possesses a flux of $\pi$, the hoppings herein cannot be all positive or negative. Figure 3(a) shows one configuration that satisfies this requirement, providing that the intra-cell hoppings $\gamma_2 = -\gamma_1$ and the inter-cell hoppings $\lambda_2 = -\lambda_1$. In particular, the unit cell exhibits a nonzero quadrupole moment if $|\lambda_{1,2}| > |\gamma_{1,2}|$ and zero quadrupole moment if $|\lambda_{1,2}| < |\gamma_{1,2}|$. These two topologically distinct quadrupole phases, denoted by **A** and **B** for brevity, can be constructed directly in acoustic systems following the above design route for positive and negative hoppings. As exhibited above, the signs of the acoustic hoppings can be controlled by the connectivity of the coupling tubes, the strengths of hoppings can be engineered by tuning the positions and sizes of the coupling tubes, and the onsite energy depends mostly by the geometry of cavity resonators. As such, topologically distinct quadrupole insulators can be achieved by tailoring the geometric parameters of the coupling tubes. In fact, the phase **B** can be simply realized by switching the values of the inter- and intra-cell hoppings predesigned for the phase **A**, associated with the same band structure. According to the generalized bulk-boundary correspondence [35,36], topologically protected in-gap states can be observed at the edges and corners formed between **A** and **B**, as to be shown in Fig. 3(d).

Figure 3(b) shows the unit structure of our acoustic QTI. It has an in-plane lattice constant of 4.2 cm and a finite height of 2.0 cm in the *z* direction. Critically, the designed acoustic structure carries positive (red) and negative (blue) hoppings, which are distributed according to the TB model in Fig. 3(a). The elementary structures are a bit more complex than those presented in Fig. 1, in order to satisfy the TB model in a quantitative way (See details in *Supplemental Materials*). Figure 3(c) shows the band structure for this acoustic system. The lower four bands (grey solid lines) are formed by the *S* modes featured with uniform field patterns inside the cavities. In contrast, the upper four bands (black solid lines) are formed exactly by $P_z$ modes, each two of which are nearly doubly-degenerate. All the $P_z$ bands are captured precisely by the TB model (green dashed lines) with fitting parameters $\lambda_1 = -\lambda_2 = -455\,\text{Hz}$, $\gamma_1 = -\gamma_2 = -57\,\text{Hz}$, and the onsite energy 5830 Hz. Remarkably, a big omnidirectional band gap (~1060 Hz, associated with a gap-to-midgap ratio ~18%) emerges between the two pairs of $P_z$ bands, which is induced by the strong dimerization ($|\lambda_{1,2}|/|\gamma_{1,2}| \approx 8.0$) and matches well the value $2\sqrt{2}(|\lambda_{1,2}| - |\gamma_{1,2}|)$ derived from the TB model (*Supplemental Materials*). In order to demonstrate the higher-order topological effect, in Fig. 3(d) we present the simulated eigenfrequency spectrum for a finite-sized sample (12×12 unit cells in total, see inset). As expected, it shows clearly the coexistence of the topologically protected 1D edge states and 0D corner states inside the bulk gap. Physically, both the topological in-gap states are originated in the bulk topology of the QTI: the gapped edge states stem from the nontrivial surface dipole quantization, whereas the topological corner states come from the nontrivial quadrupole quantization [35,36]. (As such, these in-gap topological responses depend mostly on the geometries of the coupling tubes.) In particular, the big bulk gap and the spectrally well-isolated topological corner states greatly facilitate the experimental characterization of the cascade hierarchy of quadrupole topology in our system. Note that the corner modes are not perfectly degenerate (associated with a frequency split ~1.7% with respect to the bulk gap), because of the unavoidable next-nearest-neighbor coupling or coupling to other bands [37, 55], which is beyond the description of the TB model presented in Refs. [35, 36].

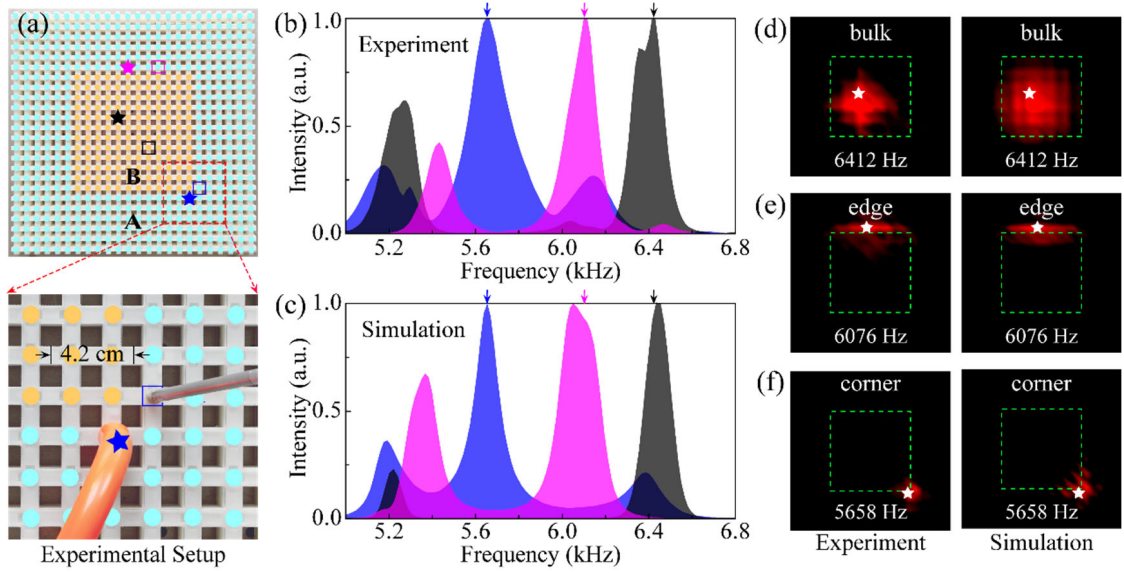

FIG. 4. (a) Experimental setup for identifying the hierarchy quadrupole topology. Upper panel: experimental sample. Lower panel: a zoomed in picture for demonstrating the pump-probe measurement of the corner response. The color stars and squares label respectively the positions of the source and detector when measuring the bulk (black), edge (purple) and corner (blue) responses. (b) Experimentally measured bulk (black), edge (purple) and corner (blue) spectra. (c) Numerical comparison of (b). (d)-(f) Measured and simulated pressure amplitude distributions for the bulk, edge and corner states, respectively, plotted in logarithmic scale with bright (dark) for strong (weak) field. The frequencies correspond to the peaks labeled in (b) and (c). In each case, the white star indicates the position of the acoustic source.

*Experimental demonstration of the hierarchy quadrupole topology.* The higher-order topology of the acoustic QTI was identified by our experiments. Figure 4(a) shows the experimental setup. As considered in Fig. 3(d), our sample has a size of $12\times12$ unit cells (576 resonators in total), where the site cavities were colored to distinguish the topologically different quadrupole phases **A** and **B**. To map out the 2D field profiles, each site cavity was perforated with a hole (sealed when not in use) for inserting the acoustic source or detector. To confirm the coexistence of the bulk, edge and corner modes, we first measured the frequency-resolved acoustic responses for three typical pump-probe configurations. In each case, the positions of the acoustic source and detector are highlighted by the color star and square [Fig. 4(a)]. Figure 4(b) shows the corresponding intensity spectra detected for the three independent experiments. As expected, the bulk spectrum (black) exhibits two intensity peaks (centered at 5260 Hz and 6412 Hz) and identifies the wide bulk gap between the two pairs of $P_z$ bands in Fig. 3(c). The edge spectrum (purple) exhibits two peaks (around 5410 Hz and 6076 Hz) within the bulk gap, which correspond exactly to the gapped 1D edge states. By contrast, in addition to the minor peaks contributed from the bulk and edge states, the corner spectrum (blue) shows one prominent peak (centered at 5658 Hz) inside the common gap of the edge and bulk states, which serves as an unambiguous evidence for the presence of topological corner states. (The peak broadening mainly comes from the unavoidable material absorption and fabrication error of the sample.) All the spectra match excellently the simulation results presented in Fig. 4(c), in which a uniform dispassion loss was used (*Supplemental Materials*). To conclusively verify the cascade hierarchy of high-order topology, we further scanned and characterized the field distributions at the peak frequencies of the bulk, edge and corner spectra. As shown in Figs. 4(d)-4(f), the measured field patterns (left panels) directly visualize the features of the bulk, edge and corner states, respectively, again in good agreement with the numerical results (right panels).

*Conclusions*. We have designed an acoustic QTI that follows the TB model proposed for QTI theory, and experimentally validated the hierarchy quadrupole topology through measuring the topological 1D edge states and 0D corner states. The positive and negative hoppings are conceived independently by introducing different connectivity between acoustic resonators. Instead of multipole resonators [37], here dipole resonators are used to achieve a wide bulk gap, which facilitates the experimental characterization of the hierarchy topology of our acoustic QTI. In addition, an arbitrary controlling of the real-valued hoppings enables the further investigation of rich physics inherent in engineering $Z_2$-gauge flux [54,56,57]. Note that the corner states are topologically stable even in the presence of disorders (see *Supplementary Materials*), as long as the chiral symmetry of the acoustic structure is ensured by its fabrication. This enables a robust and precise control of sound energy concentration, and thus points to a wide range of application avenues that demand highly-localized strong sound fields, such as sound energy-harvesting, acoustic sensing, and trapping microparticles by acoustic radiation force.

**Note added:** At the review stage of this Letter, we are aware of one work that realized acoustic QTIs based on the nonsymmorphic symmetry of sonic crystals [58]; we also notice that similar approaches were introduced to construct acoustic octupole topological insulators [59,60].

**Acknowledgements**

C. Q. is supported by the Young Top-notch Talent for Ten Thousand Talent Program (2019-2022) and National Natural Science Foundation of China (Grant No. 11774275); M. X. is supported by the National Natural Science Foundation of China (Grant No. 11904264) and the startup funding of Wuhan University; Z. L. is supported by the National Key R&D Program of China (Grant No. 2018YFA0305800) and the National Natural Science Foundation of China (Grant Nos. 11890701 and 11674250).

# Supplemental Materials for

# "Acoustic realization of quadrupole topological insulators"

Yajuan Qi[1], Chunyin Qiu[1*], Meng Xiao[1*], Hailong He, Manzhu Ke, and Zhengyou Liu[1,2]

[1]Key Laboratory of Artificial Micro- and Nano-Structures of Ministry of Education and School of Physics and Technology, Wuhan University, Wuhan 430072, China

[2]Institute for Advanced Studies, Wuhan University, Wuhan 430072, China

*Corresponding authors: cyqiu@whu.edu.cn, phmxiao@whu.edu.cn

# 1. Supplementary information for our simulations, experiments and TB model

*Numerical simulations.* All full-wave simulations were performed using a commercial solver package (COMSOL Multiphysics). The photosensitive resin material used for fabricating samples is modeled as acoustically rigid in the airborne sound environment. The air density $\rho_0 = 1.29 \, \text{kg/m}^3$ and the sound speed $c_0 = 340 \, \text{m/s}$ were used to solve the eigenproblems in Fig. 1 and Fig. 3. To capture the unavoidable dispassion loss in real experiments, an imaginary part of the sound speed $c' = 4.0 \, \text{m/s}$ was added to simulate the acoustic responses in Fig. 2 and Fig. 4. The presence of dissipation broadens the intensity peaks in the frequency spectra Figs. 2(c) and 2(d), and Figs. 4(b) and 4(c).

*Experimental measurements.* Our experiments were performed for airborne sound at audible frequency. All the samples involved in our experiments were fabricated via 3D printing using photosensitive resin material, associated with a nominal fabrication error ~0.02 cm. In order to excite and detect the desired acoustic responses, holes of diameter ~0.8 cm were perforated on the samples, which were sealed except for those used for excitation and detection. A broadband sound signal was launched from a soft tube of inner diameter ~0.8 cm, which behaved as a point source for the typical wavelength ~20.0 cm and ~6.0 cm in Fig. 2 and Fig. 4, respectively. The pressure field was recorded with a 1/4 inch microphone (B&K Type 4187) inserted inside the sample. Both the amplitude and phase of the pressure field were recorded and frequency-resolved with a multi-analyzer system (B&K Type 3560B).

*Tight-binding model and fitting parameters.* Our acoustic structure in Fig. 3(b) can be well captured by the square-lattice TB model proposed for QTI theory, in which the effective Hamiltonian can be written as

$$\mathrm{H}\left(\vec{k}\right) = \begin{pmatrix} \omega_0 & h_{12} & 0 & h_{14} \\ h_{12}^* & \omega_0 & h_{23} & 0 \\ 0 & h_{23}^* & \omega_0 & h_{34} \\ h_{14}^* & 0 & h_{34}^* & \omega_0 \end{pmatrix}, \tag{1}$$

with $h_{12} = \gamma + \lambda e^{ikx}$, $h_{14} = \gamma + \lambda e^{-iky}$, $h_{23} = h_{14}$, $h_{34} = -h_{12}^*$, where $\gamma = \gamma_1 = -\gamma_2$ and $\lambda = \lambda_1 = -\lambda_2$ are hopping strengths involved in this TB model, and $\omega_0$ is the associated on-site energy. The eigenenergies are

$$\omega_{1,2}(k_x, k_y) = \omega_0 - \sqrt{\frac{1}{2}\left[\gamma^2 + \lambda^2 + \gamma\lambda(\cos k_x + \cos k_y)\right]}, \tag{2}$$

$$\omega_{3,4}(k_x, k_y) = \omega_0 + \sqrt{\frac{1}{2}\left[\gamma^2 + \lambda^2 + \gamma\lambda(\cos k_x + \cos k_y)\right]}, \tag{3}$$

which form two pairs of doubly degenerate $P_z$ bands for the dimerized hopping strengths $\gamma \neq \lambda$. Specifically, the complete gap determined by the minimal value of $\Delta = \omega_{3,4} - \omega_{1,2}$ occurs at the M point of the Brillouin zone. It is $\Delta_{\min} = 2\sqrt{2}(\gamma - \lambda)$, which provides us a recipe to design a big band gap. As shown in Fig. 3(c), the TB model captures precisely the four $P_z$ bands obtained by our full-wave simulation, associated with fitting parameters $\lambda = 455$ Hz, $\gamma = 57$ Hz, and $\omega_0 = 5830$ Hz.

*Details for the disordered systems.* In Fig. S5 we present a theoretical discussion on the topological robustness of the corner states against the hopping disorders within a tight-binding model. The finite-size sample here contains $24 \times 24$ unit cells, with $12 \times 12$ unit cells of crystal **B** ($|\lambda_{1,2}| < |\gamma_{1,2}|$) surrounded with crystal **A** ($|\lambda_{1,2}| > |\gamma_{1,2}|$). For simplicity, we use (m, n) with $\{m, n\} \in \{1, 2, \cdots, 24\}$ to label the position of each unit cell. Thus $\{m, n\} \in \{7, 8, \cdots, 18\}$ for crystal **B** and otherwise crystal **A**. Specifically, in Fig. S5(a) the strong and weak couplings are set as 1.8 and 0.2, respectively, and in Fig. S5(b) the strong and weak couplings are set as $2 + 0.8\sqrt{2}$ and $1-0.4\sqrt{2}$. The signs of hopping in Fig. S5(a) are set according to Fig. 3(a), and the sign of hopping are all positive in Fig. S5(b). These setting results in an identical bulk gap in which the corner states embedded for both cases, $E_{gap} = 3.2\sqrt{2}$. The disorder here is applied either on the hopping strength or onsite energy with $\kappa = \kappa_0 + X\delta_\kappa E_{gap}$ or $\varepsilon = X\delta_\omega E_{gap}$, respectively. Wherein $\kappa$ ($\kappa_0$) represents the hopping coefficient with (without) disorder, $X \in [-0.5, 0.5]$ denotes a uniform random number, and $\delta_{\kappa,\omega}$ gives the strength of disorder with respect to $E_{gap}$. For each $\delta_{\kappa,\omega}$, the energies and eigenvectors of all the eigenstates are calculated and then averaged over 30 samples. The gray region is given by $G = \bigcup_i [\bar{E}_i - \sigma_i, \bar{E}_i + \sigma_i]$, where the union is over all the eigenstates, and $\bar{E}_i$ and $\sigma_i$ are the average eigenvalue and its standard derivation of the $i$-th eigenstate, respectively. We also define a projection operator to characterize whether an eigenstate is localized around the corner or not. The value of this function is given by $P = \sum_{\{m,n\}} |P_{m,n} \bar{u}_{m,n}|^2$, where $\bar{u}_{m,n}$ represents the ensemble averaged eigenvector at lattice labeled by (m, n), $P_{m,n} = 1$ when the lattice is at the corners, i.e., $\{m,n\} \in \{6, 7, 12, 13\}$, and $P_{m,n} = 0$ otherwise. The value of $P$ is then represented by the color of each eigenstates in Fig. S5, where blue (red) represents the maximum (minimal) value of $P$.

## II. Additional data and discussions

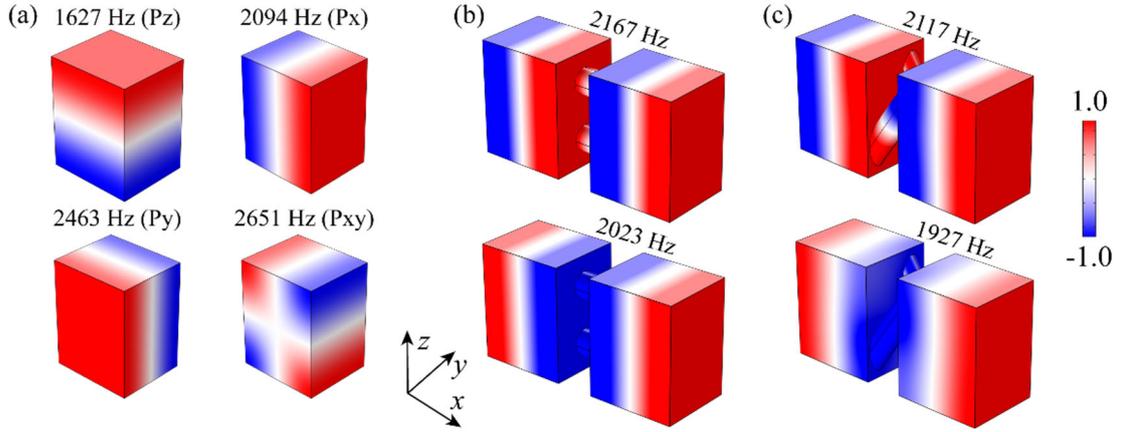

FIG. S1. (a) Low-frequency eigenmodes for the single cavity. (b) and (c): Coupled $P_x$ modes for the straight-linked and cross-linked cavity structures, respectively.

In the main text we focus on the $P_z$ modes. We have shown that the sign of hopping can be controlled by the connectivity of the double cavities: the straight connection gives a negative hopping and the cross connection offers a positive hopping, respectively. In addition to the $P_z$ mode, the cavity also supports other resonance modes at the higher frequencies, as exhibited in Fig. S1(a). However, as exemplified in Figs. S1(b) and S1(c) by the coupled $P_x$ modes, both the straight-linked and cross-linked structures exhibit positive hoppings since the in-phase coupled modes occur at the higher frequencies than those out-of-phase ones. In fact, the sign of hopping can be flipped by considering other connectivity, according to the eigenfield pattern of the cavity mode.

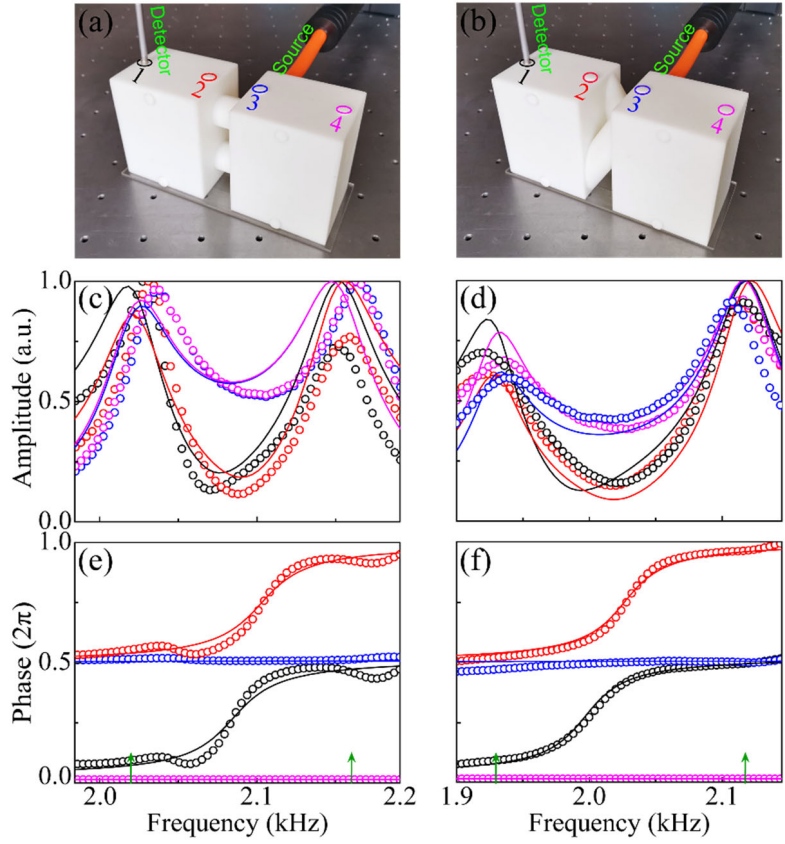

FIG. S2. Experimental identification of the positive hoppings for the coupled $P_x$ modes. (a) and (b): Experimental setups for the straight-linked and cross-linked double cavity systems, respectively. The acoustic signal is injected from the backside of the sample, and four typical positions (labeled with color circles) on the topside are used for detection. (c) and (d): Measured (circles) and simulated (lines) pressure amplitude responses at the four locations (plotted with consistent color). The green arrows indicate the eigenfrequencies of the coupled modes. (e) and (f): The associated phase spectra.

Experiments were performed to identify the positive hoppings for $P_x$ modes under the current connectivity. Different from the measurements for $P_z$ modes (see Fig. 2), here the acoustic responses of four typical positions on the topside of the samples were measured [see Figs. S2(a) and S2(b)]. As shown in Figs. S2(e) and S2(f), the phase spectra for both structures demonstrate qualitatively the same behavior. First, there is a phase difference $\sim \pi$ between the positions 1 and 2 (or 3 and 4), as an indication of the $P_x$ cavity modes. Second, the phase difference between the positions 1 and 3 (or 2 and 4) approaches $\pi$ at the lower frequency while approaches zero at the higher frequency. This concludes positive hoppings for both configurations, since the in-phase coupled modes both occur at the higher frequency.

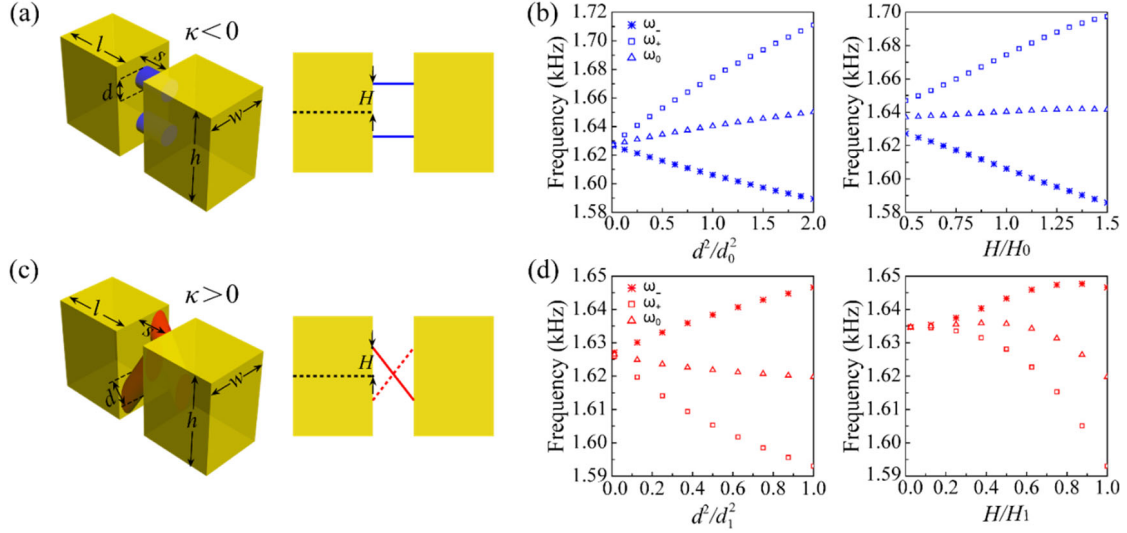

FIG. S3. Parameter dependences of the split eigenfrequencies $\omega_\pm$ and associated central frequency $\omega_0$ for the two double-cavity structures involved in the main text. (a) Left: 3D plot of the straight-linked structure. Right: A front view that sketches the position of the tube axis with $H$. (b) Left: The quantities $\omega_\pm$ and $\omega_0$ plotted as a function of $d^2/d_0^2$, the dimensionless cross section area of the narrow tubes. Right: The quantities $\omega_\pm$ and $\omega_0$ plotted as a function of $H/H_0$, the dimensionless distance between the tube and the middle *xy*-plane of the cavity (associated with zero pressure). Here $d_0 = 2.32$ cm and $H_0 = 2.61$ cm are the values used in our main text. (c) and (d): Similar to (a) and (b), respectively, but for the cross-linked structure. Here $d_1 = 2.32$ cm and $H_1 = 2.90$ cm are the values used in our main text.

Now we consider the parameter dependences of the split eigenfrequencies $\omega_\pm$ and associated central frequency $\omega_0$ for the coupled $P_z$ modes. In both cases, our numerical results show an increasing hopping strength $|\kappa|=|\omega_+ - \omega_-|/2$ with the growth of the tube size ($d^2$) or the distance ($H$) to the zero pressure plane, because of the stronger and stronger wavefunction overlapping between the two cavities. Note that the central frequency $\omega_0 = (\omega_+ + \omega_-)/2$ may shift when changing these parameters.

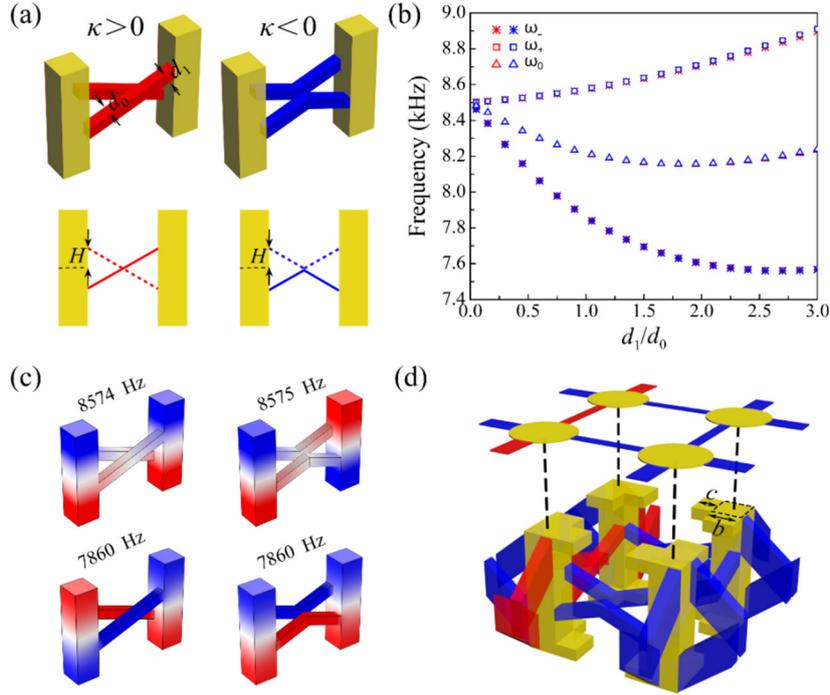

FIG. S4. Structure design of the unit cell. (a) Upper panels: 3D plots of the double-cavity structures with positive (left) and negative (right) hoppings. Unlike the straight-linked structure, here the V-shape tubes are used for realizing the negative hopping. Lower panels: Cross section views that sketch the positions of the tube axes with the parameter $H$. (b) The split eigenfrequencies $\omega_\pm$ and central frequency $\omega_0$ plotted as a function of the tube width $d_1$ for both the cross-linked and V-shape-linked structures, in which the other geometry parameters are fixed. (c) Eigenfield patterns for these two structures, exemplified with $d_0 = d_1 = 0.20$ cm. (d) Unit structure of the acoustic QTI used in our main text.

As stated above, in addition to the hopping sign determined by the connectivity, the coupling strength and the central frequency will also be influenced by the geometric details of the coupling tubes. This leads to a great challenge to construct an acoustic structure that quantitatively satisfies the TB model: $\gamma_1 = -\gamma_2$ and $\lambda_1 = -\lambda_2$, associated with a universal central frequency for these four elementary structures. Here we conceive a simple approach to relax the dilemma. As shown in Fig. S4(a), we start from the structure with positive coupling, i.e., cross-linked double cavities. (The cavity parameters and the distance between the cavities are fixed as $l = 0.50$ cm, $w = 0.50$ cm, $h = 2.0$ cm, and $s = 1.60$ cm.) Unlike the straight-linked structure used before, here the tubes are intentionally bent into V-shape to achieve a negative hopping. This artful design guarantees an identical central frequency and coupling strength to the cross-linked structure, as long as we fix the size of the coupling tubes (characterized with thickness $d_0$ and width $d_1$) and their positions linking to the cavities (characterized with $H$). This can be seen clearly in Fig. S4(b), the split eigenfrequencies and central frequency plotted as a function of the tube width $d_1$ for both the cross-linked and V-shape-linked elementary structures, in which $d_0 = 0.20$ cm and $H = 0.29$ cm are fixed. The eigenfrequencies $\omega_+$ and $\omega_-$ for the in-phase and out-of-phase coupled modes are determined by the corresponding eigenfield patterns, as exemplified in Fig. S4(c) by setting $d_1 = 0.20$ cm. Therefore, we can construct each pair of couplings $\gamma_{1,2}$ and $\lambda_{1,2}$ independently, and tune the strengths $\gamma = |\gamma_{1,2}|$ and $\lambda = |\lambda_{1,2}|$ by changing the position or size of the coupling tubes. In order to achieve a big band gap, a strong dimerization is considered by using different tube width and tube position: $d_1 = 0.20$ cm and $H = 0.29$ cm to attain the weak coupling $\gamma = |\gamma_{1,2}| = 57$ Hz, and $d_1 = 0.47$ cm and $H = 0.67$ cm to achieve the strong coupling $\lambda = |\lambda_{1,2}| = 455$ Hz, respectively. However, as exemplified in Fig. S4(b), the central frequency will shift when engineering the coupling strength. To

reduce this effect, additional side cavities are further introduced for fine-tuning. This leads to the final unit structure used our main text, as shown in Fig. S4(d), with $b = 0.50$ cm and $c = 0.37$ cm.

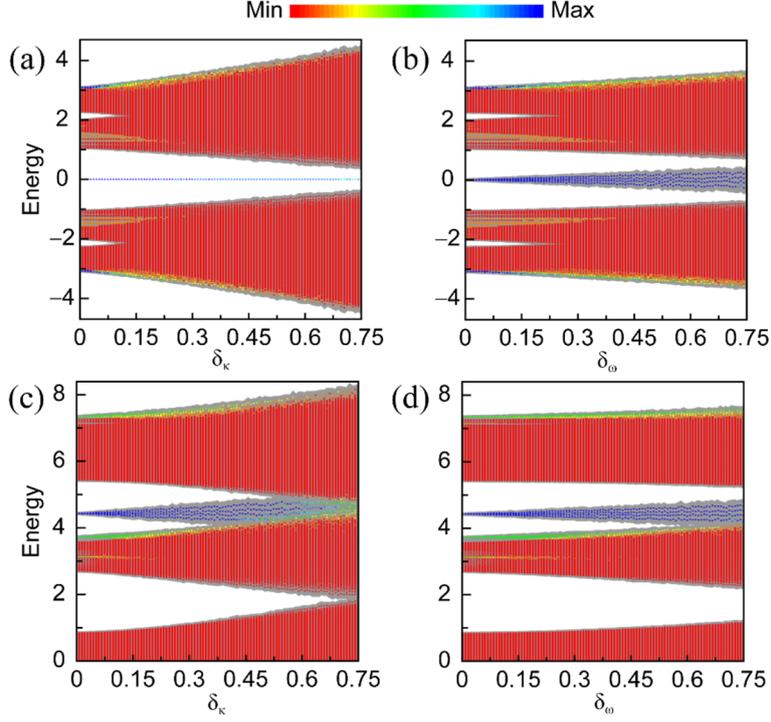

FIG. S5. Topological robustness of the corner states against disorders. (a, b) Ensemble average of the eigenenergies (colored dots) at different strength of randomness ($\delta_{\kappa,\omega}$) in our acoustic QTI. (c, d) Comparison results for a 2D higher-order topological insulator formed by all-positive hopping coefficients (See Refs. 42 and 43 in the main text). The gray region represents the standard derivation of the eigenspectra. The color of each dot characterizes whether the eigenstate is localized around the corner: Blue means an eigenstate localized around the corner and red otherwise.

Here we study the topological robustness of the corner states against the randomness on the hopping coefficients and resonance frequencies. To be more specific, we add uniformly distributed random number on the hopping coefficients and resonance frequencies with the strength of randomness given by $\delta_\kappa$ and $\delta_\omega$, respectively. (See Sec. 1, *Details for the disordered systems*). To reduce the numerical workloads, the calculation here is performed within tight-binding models, which has been demonstrated to be in good agreement with the full-wave simulations (see Fig. 3 in our main text). As a control group, we also calculate the case when all the hopping coefficients are positive, i.e., $\lambda_1 = \lambda_2 > 0$, $\gamma_1 = \gamma_2 > 0$ and $\gamma_{1,2} > 2\lambda_{1,2}$, which are also predict to possess topological corners states [42, 43]. The ensemble average of eigenspectra (colored dots) are given in Fig. S5 as functions of $\delta_\kappa$ or $\delta_\omega$, where Figs. S5(a) and S5(b) represent the spectra of QTI and Figs. S5(c) and S5(d) represent the cases when all the hopping coefficients are positive. The gray background shows the standard derivation of the spectra. For comparison, here we choose the hopping strengths such that the bulk band gap wherein the corner states embedded for both cases are the same. The color here tells whether an eigenstate is localized around the corner or not: blue means that the eigenstate concentrates at the corner and red the opposite.

Without randomness, the bulk states cover energy range $[-3.1, -2.2] \cup [2.2, 3.1]$ in Figs. S5(a) and S5(b) and $[-7.4, -5.4] \cup [-0.9, 0.9] \cup [5.4, 7.4]$ in Figs. S5(c) and S5(d), and the surface states cover energy range $[-2, -1] \cup [1, 2]$ in Figs. S5(a) and S5(b) and $[-3.7, -2.7] \cup [2.7, 3.7]$ in Figs. S5(c) and S5(d). With the increasing of randomness ($\delta_{\kappa,\omega}$), bulk states and edge states gradually merge together in all subfigures in Figs. S5. The energies of the corner states in Fig. S5(a) localized around 0 while those in Fig. S5(c) spread and merge into the bulk and surface states. This can also be seen from the "color" of each eigenstates, which are blue for all the corner states in Fig. S5(a) and approach yellow in Fig. S5(c) with the increasing of $\delta_\kappa$. Thus one can conclude that, compared with the system made of all-positive hoppings, the corner states of the QTI are much more robust against the hopping disorders. On the other hand, the corner states in both Figs. S5(b) and S5(d) are not robust against randomness on the onsite potential as the chiral symmetry is broken here. With the increasing of $\delta_\omega$, the spectra of corner states in both Figs. S5(b) and S5(d) gradually spread and will merge into the edge state and bulk states for large enough randomness. Note that for small values of $\delta_\kappa$, Fig. S5(a) also shows 'corner' states emerging at the outer boundaries of the bulk states (associated with energies around +3 and -3). These states correspond to topologically trivial defect modes localizing at the corners, whose localization degrades quickly with the increase of $\delta_\kappa$. Similar phenomena also occur in Fig. S5(b)-S5(d), while appearing at different energies.

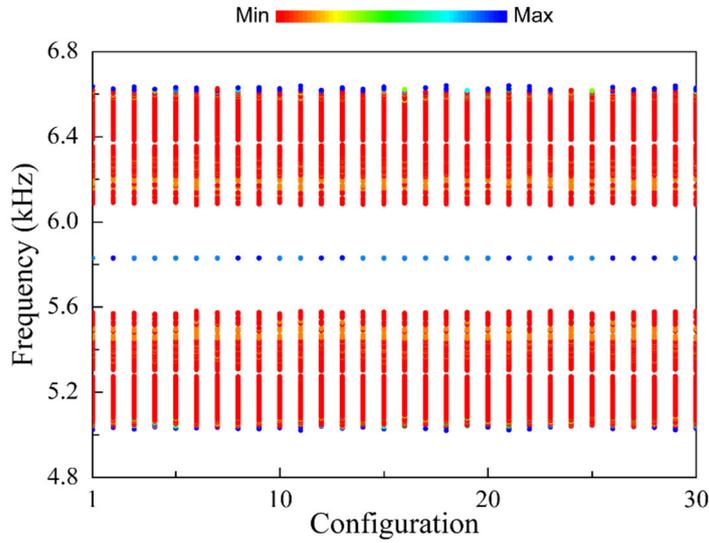

Fig. S6. Energy spectra for 30 independent random configurations, which involve randomness in hopping constant and resonant frequency simultaneously. The color of each dot characterizes whether the eigenstate is localized around the corner: Blue means an eigenstate localized around the corner and red otherwise.

Our experimental sample was fabricated by 3D printing technique, associated with a resolution of ~0.1mm. This gives an estimation of relative deviation of the hopping constant (~5%), and a relative deviation of the resonant frequency (~0.5%). Based on these parametric perturbations, we have calculated the energy spectra for 30 independent configurations. The result is presented in Fig. S6, which shows a weak energy fluctuation in the whole spectra for the 30 configurations. In particular, the corner modes are almost pinned to the middle of the band gap, since the disorders are relatively weak.